\documentclass[aps,prc,twocolumn]{revtex4}
\usepackage{graphicx,amssymb}

\begin{document}

\title{Bell Locality and the Nonlocal Character of Nature}
\author{Travis Norsen}
\affiliation{Marlboro College \\ Marlboro, VT  05344 \\ norsen@marlboro.edu}

\date{\today}

\begin{abstract}
It is demonstrated that hidden variables of a certain type 
follow logically from a certain local causality requirement (``Bell Locality'')
and the empirically well-supported predictions of 
quantum theory for the standard EPR-Bell setup.
The demonstrated hidden variables are precisely
those needed for the derivation of the Bell Inequalities.  We
thus refute the widespread view that empirical violations of Bell
Inequalities leave open a choice of whether to reject (i) locality or
(ii) hidden variables.  Both principles are indeed assumed in the 
derivation of the inequalities, but since, as we demonstrate here,
(ii) actually follows from (i), there is no choice but to blame the
violation of Bell's Inequality on (i).  Our main conclusion is thus 
that no Bell Local theory can be consistent with what is known from 
experiment about the correlations exhibited by separated particles. 
Aside from our conclusion being based on a different sense of locality
this conclusion
resembles one that has been advocated recently by H.P. Stapp.
We therefore also carefully contrast
the argument presented here to that proposed by Stapp.

\bigskip


\end{abstract}

\maketitle

\section{INTRODUCTION}
\label{sec1}

Bell's Theorem \cite{bell} has given rise to a unique situation in
science. 
Despite being hailed from all quarters as fundamentally important
(``the most profound discovery of science,'' says H. P. Stapp
\cite{stappquote}),
the \emph{meaning} of the theorem -- what, exactly, it proves -- remains
widely and hotly contested.  There are (at least) two camps.  In the 
first camp are those who believe
that Bell's Theorem proves that non-locality is a fact of nature,
which must be reflected in any adequate theory.  
\cite{bell,maudlin,shelly,norsen,wiseman}
Then there are, in the second camp,
those who believe that Bell's Theorem proves only the impossibility of  
empirically viable local ``hidden variable'' alternatives to orthodox 
quantum theory. \cite{wigner,merminquote}

It is widely accepted (by people in both camps) that two assumptions
-- locality and hidden variables -- are needed in the derivation of
the Bell Inequalities.  Those in the first camp, however, frequently
cite the EPR argument \cite{epr} as providing a link between these two
principles:  the existence of hidden variables, it is argued, 
follows from the requirement of locality.  For example, Bell writes that
\begin{quote}
``My own first paper on this subject [Bell's Theorem] ... 
starts with a summary of the EPR argument \emph{from locality to} 
deterministic hidden variables.  But the commentators have almost 
universally reported that it begins with deterministic hidden 
variables.'' \cite[pg 157]{bell}
\end{quote}
Those in the second camp have either failed to understand the role of
the EPR argument as part-one of Bell's two-part argument for
non-locality, or have (following Bohr and his supporters) rejected the
EPR argument as invalid.  

The main purpose of the present paper is to present a new version of
the EPR argument (from locality to deterministic hidden variables, the
very hidden variables needed to then derive a Bell Inequality) which
removes these objections and hence solidifies the viewpoint of the
first camp.  The 
argument to be presented differs in several important ways from
the original EPR argument.  First, instead of merely attempting to
prove the \emph{incompleteness} of the orthodox (wave-function-only)
description of quantum states, we provide an actual constructive proof
of the existence of particular hidden variables (that is, their
required existence under the
assumption of local causality).  Second, in place of EPR's somewhat
obscure locality assumption, we utilize Bell's mathematically precise
definition of local causality (hereafter ``Bell Locality'').  This
locality condition has been widely misunderstood and misrepresented in
the Bell literature, so we discuss and defend it in some detail.  The
present paper thus completes the logically rigorous demonstration that
empirical violation of Bell's Inequalities (under appropriate,
delayed-choice conditions) signals the violation in Nature of Bell
Locality.  In short, we remove the basis (objections to EPR) of the
second camp's arguments against the first camp.

Why is this debate between camps one and two even important?  
Because the life or
death of the hidden variables program (advocated by those such as
Einstein, Bohm, and Bell who were dissatisfied with the Copenhagen
quantum theory) hangs in the balance.  Those in the first camp 
argue that, since non-locality is a real fact about the
physical world, physicists should remain open to
explicitly non-local alternatives to orthodox quantum theory
such as Bohmian Mechanics.  \cite{bohmmech}
On the other hand, those in the second camp use Bell's Theorem against
the hidden variables program, by arguing that any such alternatives to
orthodox theory must conflict with relativity and hence needn't be
seriously considered.

This standard argument of the second camp against the first 
was clearly articulated by H. P. Stapp.  Contrary to the view of 
those in the first camp, he writes, Bell's Theorem ``shows only that if
certain predictions of quantum theory are correct, \emph{and if a certain
hidden-variable assumption is valid}, then a locality condition
must fail.  This locality condition expresses the physical idea,
suggested by the theory of relativity, that what an experimenter
freely chooses to measure in one spacetime region can have no effect
of any kind in a second region situated spacelike relative to the
first.''  Therefore,
``the most natural conclusion to draw [from the empirical violation of
Bell's Inequality] is not that locality fails, but
rather that the hidden-variable assumption is false.''  After all:  
``Bell's
hidden-variable assumption was an ad hoc assumption that had no
foundation in the quantum precepts.  Indeed it \emph{directly 
contradicted the quantum precepts}.''  \cite{stapp:btwhv}

This is a widely held view whose implications for the hidden-variables
debate were eloquently captured by N. David Mermin:  ``To
those for whom nonlocality is anathema, Bell's Theorem finally spells
the death of the hidden-variables program.'' \cite{merminquote}

In an interesting recent series of papers
H. P. Stapp has questioned the reasoning behind this
conclusion and presented arguments in support of the first camp. 
\cite{stapp:btwhv,stapp1,stapp2}  
Stapp's re-thinking is based on his attempt to 
prove a \emph{stronger version} of Bell's Theorem which dispenses
altogether with the assumption of hidden-variables.
(Actually, Stapp's argument is based on Hardy's scenario \cite{hardy} 
in which a simple logical contradiction -- rather than an inequality
-- is deduced from the assumption of
local hidden variables and quantum theory's probabilistic
predictions.  But this distinction is not relevant here.)  

We believe, however, that Stapp's arguments for the first-camp
position are fatally flawed.  So, by way of clarifying the content of
the first-camp-supporting arguments to be presented here, we shall
contrast these with the arguments of Stapp.  Let us then briefly
review the logical structure of Stapp's attempts, to see how his
approach differs from the approach of Bell (elaborated and completed
in the present paper).

Using a certain locality principle, the predictions of quantum 
theory for the Hardy
scenario, and an assumption about which experiment was freely chosen
in the left ($L$) wing of the experiment, Stapp establishes the truth 
of a certain statement (call it $S$) which refers to possible
measurements and their outcomes on the right ($R$). 
He then proves that, had the
experimenter in region $L$ made a different free choice, $S$ would
cease to be true.  Therefore, the truth status of a statement pertaining
exclusively to region $R$ depends on a free choice made in region $L$,
which we may assume to be space-like separated from $R$.  According to
Stapp, this constitutes a kind of non-local action at a distance, and
thus establishes that not merely empirically viable hidden-variable
theories but, rather, \emph{any} theory in agreement with quantum
theory's 
predictions, must violate locality.  It establishes, says Stapp, the
ineliminable non-local character of quantum theory itself.

Stapp's argument, however, has been roundly criticized.  The trouble
pertains to Stapp's definition of locality, which definition William 
Unruh summarizes as follows:  ``the
truth of a statement pertaining exclusively to possible events in
region $R$ cannot depend on which free choice is to be made by the
experimenter in region $L$.''  \cite{unruh}
As Mermin has clarified, this definition encompasses
two very different ``locality'' ideas:  first, that what
an experimenter on one side chooses to measure can have no influence
on the actual outcome of an actually-performed measurement on the
other side; and second, that what an experimenter on one side chooses to
measure can have no influence on what can be validly inferred about
what \emph{would} have happened had some \emph{other} choice been made
on the other side.  \cite{mermin}  Violation of the first sort of
locality would indeed appear to be in
conflict with relativity's prohibition on superluminal causation; but
it is only the latter (which seems unproblematic from the point of
view of relativistic causality) that is established by Stapp's
argument.   For Stapp's statement $S$ is not merely a statement about the
actual result of an actual experiment in $R$; it is rather a
counterfactual conditional asserting that a certain outcome
\emph{would} have been obtained had a different measurement been made,
given that, in fact, the actual experiment was performed and had a
certain particular outcome.  

The main objection of Mermin and others is not
simply that the statement $S$ is a counter-factual conditional and
hence not directly observable.  The critics accept the validity in
principle of such counterfactual statements, but point out that, 
although the counterfactual statement $S$ may appear superficially to
``pertain exclusively to region $R$'', it nevertheless refers
\emph{implicitly} to region $L$ via the fact that our grounds for
accepting $S$ include statements explicitly pertaining to $L$.  
(The precise
sense in which the meaning of $S$ includes implicit
reference to region $L$ is
made particularly clear in the essays of Shimony and 
Stein. \cite{shimony,shimonystein} )    As Mermin explains,
Stapp's conclusion ``does not mean ... that a choice
  made [on the left] can influence events [on the right].  What it
  does mean is that \emph{a choice made [on the left] can influence
  statements about events [on the right] that might have happened but
  did not}....  The choice of experiment on the left cannot affect
  \emph{what actually happened} ... on the right.  But it can affect
  \emph{the kinds of inferences} one can make about hypothetical
  behavior on the right.'' \cite{mermin}  And elsewhere:  ``because the 
  influence [which Stapp's argument actually establishes] is only on 
  the possibility of satisfying Stapp's criterion for the valid use 
  of counterfactuals, it is inappropriately characterized as ... 
  `non-local'.'' \cite{mermin2}  

We believe the various criticisms of Stapp's arguments are correct.
But as already indicated, we also believe that Stapp's conclusion is correct:  
empirical violation of
Bell's inequalities demonstrates the non-locality of \emph{Nature}, not
merely of viable hidden-variable theories.  Thus, another way to frame
the goal of the present paper is this:  we give an alternative to
Stapp's (failed) attempt to prove this shared conclusion.  

The argument to be presented (and it is really just a detailed 
fleshing-out of the argument hinted at by Bell)
is different from Stapp's in several
important ways.  First, instead of trying to arrive at an empirically
testable inequality (or a logical contradiction) without 
assuming the existence of hidden
variables, we show that hidden variables (of a certain specific
character to be identified) \emph{must exist} under the assumption of
locality.  Then, it is non-controversial to point out that Bell's
Theorem (in its original form) renders any hidden-variables theory
of this type
inconsistent with experiment.  Our argument is thus not a different
(``stronger'') \emph{version} of Bell's Theorem; it is not a ``Bell's
Theorem without hidden-variables.'' \cite{bellwohv}  
Rather, it is a proof that the hidden-variable and locality
assumptions needed to derive Bell's Inequality are not on an equal
logical footing in the overall argument.
The hidden variables posited by Bell are 
not an ``ad hoc assumption'' but, rather, a logical implication 
of locality.

Second, our locality criterion differs from the various
senses of locality utilized by Stapp.  As mentioned, we present and 
then use the straightforward mathematical local
causality condition introduced and advocated by J. S. Bell:  ``Bell
Locality.''  

Third, where Stapp (and his critics) focus on specific (possible)
measurement outcomes, our emphasis is instead on \emph{theories}.  In
particular, the question from which our derivation proceeds is:  What
properties must a theory possess if it is to satisfy Bell Locality and
agree with certain empirical facts?  This emphasis on the structure of
theories will be seen to follow in a natural way from
Bell's locality condition, elaborated in Section \ref{sec2}.

The upshot of these differences is that our proof is immune to the
kinds of objections given against Stapp by (for example) Mermin, 
Unruh, Shimony and Stein.  The argument
presented here thus supplements Bell's Theorem in just the way
needed to conclusively establish the following (Stapp-like)
conclusion:  \emph{no Bell Local theory can be consistent with the 
predictions of quantum mechanics, i.e., with the empirical facts}.  
Note in particular that the conclusion is not
restricted to hidden-variable theories, deterministic theories, or any
other class.  It applies to all theories, orthodox quantum theory very
much included.  Finally, given that no Bell Local theory can be
empirically viable, it follows that Nature does not respect
Bell Locality.  This is the conclusion that is meant to be conveyed by
the paper's title.

\section{BELL LOCALITY}
\label{sec2}

Perhaps the single greatest source of confusion over what, exactly,
Bell's Theorem proves is confusion and ambiguity over the meaning of
``locality.''  Many different senses of locality have been defined in the
literature. \cite{maudlin,eberhard}  Our purpose here is not to review 
this literature, nor to
present a conclusive argument for the appropriateness of any
specific definition.  Rather,
we will simply pick a particular definition of locality and show that
an interesting conclusion follows from it.

The one in question is (what we shall call) Bell Locality.  This is
essentially the locality criterion that was used, ever since his
initial 1964 paper, in Bell's derivations of the famous inequality.  
His most detailed discussion of this locality concept may be found
in the article ``La Nouvelle Cuisine.'' \cite[pages 232-48]{bell}
Here we shall simply state the condition and then make a few clarifying 
remarks.  

The following is Bell's description of the locality condition, along with his
accompanying figure:
\begin{quote}
``A theory will be said to be locally causal [i.e., what we are calling
Bell Local] if the probabilities attached to values of local beables
in a space-time region 1 are unaltered by specification of values of
local beables in a space-like separated region 2, when what happens in
the backward light cone of 1 is already sufficiently specified, for
example by a full specification of local beables in a space-time
region 3.''  \cite[page 240]{bell}
\end{quote}

\begin{figure}[h]
\begin{center}
\includegraphics[width=3.0in,clip]{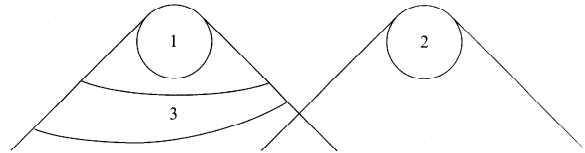}
\end{center}
\caption{
\label{fig1}
``Full specification of what happens in 3 makes events in 2 irrelevant
for predictions about 1 in a locally causal [i.e., Bell Local]
theory.''  (Figure and caption are from \cite[pg. 240]{bell}.)
}
\end{figure}

As Bell explains, ``It is important that region 3
completely shields off from 1 the overlap of the backward light cones
of 1 and 2.  Otherwise the traces in region 2 of causes of events in 1
could well supplement whatever else was being used for calculating
probabilities about 1.  The hypothesis is that any such information
about 2 becomes redundant when 3 is specified 
completely.''\cite[pg. 240]{bell}  

Let us now apply this to the standard EPR-Bell setup
in which a particle source emits
pairs of oppositely-directed neutral spin-1/2 particles in (what
orthodox quantum theory would describe as) a spin singlet state.  Two
experimenters, Alice and Bob, are located at some distance from the
source; each possesses a Stern-Gerlach type device which allows them
to measure the component of the spin of their particles (along,
respectively, the directions $\hat{a}$ and $\hat{b}$), yielding
outcomes $A = \pm 1$ and $B= \pm 1$.  Note that we consider $\hat{a}$
and $\hat{b}$ to be free variables -- not in any way affecting,
dependent on, or
correlated with one another or with the prior state of the
particle pair, whose \emph{complete description} on the space-like
hypersurface indicated in Figure 2 we denote $\lambda$.  (In
principle, a theory might assert that the outcome of the experiments
depends not just on the settings $\hat{a}$ and $\hat{b}$ and the state
of the particle pair, but also on some other facts about the measuring
apparatus; we may always include such facts in our definition of
$\lambda$ without affecting the argument.  Thus, throughout the
following, the phrase ``state of the particle pair'' should be read as
``state of the particle pair, detection apparatuses, and anything else
-- excluding of course $\hat{a}$ and $\hat{b}$ -- 
upon which the theory in question claims the outcomes depend.'')
We shall also assume that the space-time regions of Alice's and Bob's
measurements (indicated by the circles in Figure 2) are space-like
separated (as shown in the Figure).

\begin{figure}
\begin{center}
\includegraphics[width=3.0in,clip]{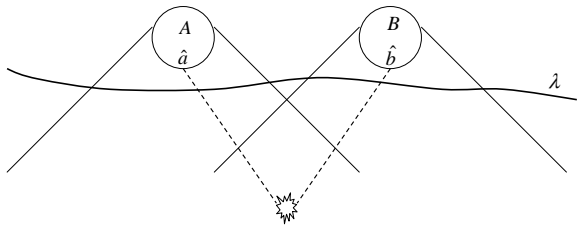}
\end{center}
\caption{
\label{fig2}
A pair of neutral spin-1/2 particles
is created, with the individual particles heading
to the left and right, where Alice and Bob freely choose an axis
(respectively $\hat{a}$ and $\hat{b}$) along which to measure their
particle's spin component.  The outcomes of those experiments are
given by $A = \pm 1$ and $B = \pm 1$.  The symbol $\lambda$ denotes a
complete specification of the state of the particle pair (and, if
necessary, the pre-measurement state of the detection apparatuses) across
some space-like hypersurface satisfying (for both measurement events)
the constraints Bell describes for his region 3 from Figure 1.
}
\end{figure}

Bell Locality then entails the following:  once we specify a
complete description of the pre-measurement state of the particle 
pair, the probability for Alice to obtain a certain outcome $A$ for a
measurement along a certain direction $\hat{a}$ is \emph{independent}
of the setting ($\hat{b}$) and outcome ($B$) of Bob's experiment.  In
particular, the probability in question does not change depending on
whether we do or do not specify this information about Bob's
experiment.  In Bell's words, that information about the
distant experiment ``becomes redundant'' when the state of the 
particle pair is specified completely.
Formally:
\begin{equation}
P(A|\hat{a},\hat{b},B,\lambda) = P(A|\hat{a}, \lambda)
\label{bellloc}
\end{equation}
where $P(X|Y)$ means the conditional probability of $X$ given $Y$.
Bell Locality of course also requires the same condition for Bob's
outcomes:
\begin{equation}
P(B|\hat{a},\hat{b},A,\lambda) = P(B|\hat{b},\lambda).
\label{belllocb}
\end{equation}
There are several key points here:
\begin{itemize}

\item The symbol $\lambda$ appearing in the definition of Bell
  Locality refers to a \emph{complete} description of the state of the
  particle pair (and, as needed, facts about the experimental apparatuses
  -- other than the freely-choosable settings -- on which the probabilities may
  depend) along the space-like hypersurface shown in Figure 2.  

\item  In
  principle, only some particular candidate theory can tell us what
  $\lambda$ consists of.  The probabilities about which Bell Locality
  speaks thus refer
  not to empirical frequencies of given outcomes, but rather to
  probabilities \emph{as assigned by the fundamental dynamics of 
  some particular theory} -- namely, whatever theory posits $\lambda$
  as a complete state description.  Thus, in the
  primary sense, what Bell Locality constrains is \emph{theories}.  
Bell himself stressed this point in ``Free variables and local
  causality'' \cite[pg 100-104]{bell}.  Here is the relevant passage:  
``I would insist here on the 
distinction between analyzing various physical theories, on the one 
hand, and philosophising about the unique real world on the other
hand.  In this matter of causality it is a great inconvenience that
the real world is given to us once only.  We cannot know what would
have happened if something had been different.  We cannot repeat an
experiment changing just one variable; the hands of the clock will
have moved, and the moons of Jupiter.  Physical theories are more
amenable in this respect.  We can \emph{calculate} the consequences of
changing free elements in a theory, be they only initial conditions,
and so can explore the causal structure of the theory.  I insist that
[my local causality criterion] is primarily an analysis of certain kinds 
of physical theory.''  Note also that Bell's characterization of the
  locality condition (the first block-quote in this section) 
  begins:  ``A \emph{theory} will be
  said to be locally causal if...'' [emphasis added].
  Given some candidate theory, we may ask:  Does this theory
  respect Eq. (\ref{bellloc}), or not?  That is:  Is this theory
  Bell Local, or not?  This is the source of the theory-emphasis (as
  opposed to experimental-outcome-emphasis) mentioned in the
  introduction.  

\item  Given this need to focus on theories, it is necessary to
  clarify what we mean by a theory.  We will stipulate just
  what is required to permit an unambiguous determination of whether or
  not a theory satisfies Bell Locality.  Thus:  a theory must
  provide some particular account of $\lambda$.  It must tell us what
  a complete description of the state of the particle pair consists
  of.  And it must provide, on that basis, 
  a definite formal structure by which the
  probabilities for the relevant possible experimental outcomes can be
  calculated.  

\end{itemize}

One important logical implication of Bell Locality is that it requires
joint probabilities for space-like separated events to
\emph{factorize} when $\lambda$ is specified:
\begin{eqnarray}
P(A,B|\hat{a},\hat{b},\lambda) &=& P(A|B,\hat{a},\hat{b},\lambda)
\times P(B|\hat{a},\hat{b},\lambda) \nonumber \\
&=& P(A|\hat{a},\lambda) \times P(B|\hat{b},\lambda).
\label{factorize}
\end{eqnarray}
The first line merely expresses the joint probability in terms of a
conditional probability, while the move to the second line involves a
trivial application of Eqs. (\ref{bellloc}) and (\ref{belllocb}).  
As Bell remarks, ``Very often
such factorizability is taken as the starting point of the analysis.
Here we have preferred to see it not as the \emph{formulation} of
`local causality', but as a consequence thereof.'' \cite[pg 243]{bell}

In principle, nothing more about Bell Locality need be said.  It is a
straightforward mathematical condition that permits an unambiguous
answer to the question:  Is Theory X Bell Local?  However, some
additional remarks may help clarify the condition and motivate the
subsequent derivation.

Bell intended this mathematical locality condition as a precise
statement of relativity's prohibition on superluminal causation.  The
idea here is that the causes of a given event should be located 
exclusively in that event's past light cone -- i.e., a complete 
specification of physical states in the past light cone should be
sufficient to uniquely and finally define the probability of the
event in question (such that specification of additional information
from a space-like separated region is redundant) -- i.e., the
probability \emph{attributed} by the theory in question to that event
should depend only on facts in the event's past light cone (and not on
facts outside the past light cone).  As Bell discusses in ``The theory
of local beables'' \cite[pg 52-62]{bell} this is a natural extension
of the obvious definition of local causality for \emph{deterministic}
theories (namely, the particular outcome assigned by the theory should
depend only on facts in the past light cone) to theories which are
not deterministic, which are irreducibly stochastic.  Such theories
replace definite predictions for which outcome will occur, with
definite predictions for the \emph{probabilities} of various
\emph{possible} outcomes.  Bell Locality is then simply the same
requirement for those \emph{probabilities} (namely, that they not depend on
facts outside the past light cone) that the obvious definition of
locality for deterministic theories imposes on those theories' outcome
predictions.  

Since there has been such confusion about this in the literature, let
us elaborate a bit more.  Any causal dependence of a given event
on other events outside the past light cone would involve superluminal
(or reverse-temporal) influence -- something that is supposed to be
ruled out by the space-time structure of relativity (and common-sense
ideas about the meaning of cause and effect).  Bell's idea was
that the assertion of such stochastic dependence by a \emph{theory} 
is tantamount to the assertion of a \emph{causal} dependence (recall
Bell's comments in the second bulleted point above).
And so relativity's prohibition on superluminal causation
requires that the probability assigned, by a theory, 
to a given event (Alice's
outcome, say) shouldn't depend (according to the postulated dynamics
of that theory) 
on either the setting of Bob's apparatus ($\hat{b}$)
or on the outcome of his experiment ($B$) or anything else outside the
past light cone of $A$.  As Bell explains, such
information should (for a relativistic theory) be \emph{redundant} (or
just irrelevant).  

According to relativity, information pertaining to a
space-like separated region ought to be
\emph{causally irrelevant} to the event in question.  This does not
necessarily mean, however, that such information must be statistically
irrelevant at the level of relative frequencies of outcomes.   
Correlations between space-like separated events which are not directly
causally related is certainly possible.
\footnote{Note that this is 
simply the denial of the contrapositive of the well-known fallacy:
``correlation implies causality.''  If correlation does not imply a 
direct causal relation between two variables, then neither does the
absence of a direct causal relation imply that the variables must be
uncorrelated.}  
They could be
correlated, for example, by virtue of their being both effects of some 
other, earlier shared cause.  But such an earlier shared cause would 
evidently have to be included in our (by hypothesis) 
\emph{complete} specification of
the state of the particle pair, $\lambda$.  For a given event (such 
as $A$),
$\lambda$ already includes all the information (other than $\hat{a}$)
that is (according to the theory in question) causally
relevant to that event.  And therefore, the probability assigned to
that event under condition $\lambda$, will not (in a Bell Local theory)
depend on additional information pertaining to a space-like
separated region.  The event in question may turn out to be
correlated with some such additional information at the level of
relative frequencies, but its probability should not depend on that
information at the level of the fundamental theoretical dynamics.

To all appearances, Bell Locality seems
to be exactly what is required by relativity's prohibition on
superluminal causation.  Any theory which violates Bell Locality 
necessarily posits a causal influence between space-like separated
regions of space-time.

There is no real point in claiming that Bell Locality is the only
possible reasonable mathematical specification of relativistic
causality.  We wish to stress,
however, that Bell Locality is \emph{prima facie} reasonable
in this role.
There is excellent reason to think that a genuinely relativistic
theory must respect Bell Locality.
From the standard point of view according to which relativity theory
is accepted as an established fact, it would be a 
\emph{surprise} to find that Bell Locality could not be maintained.
Yet this is just what Bell's two-part argument establishes.

Before moving on to the proof of the first part of this argument, 
let us stress here several 
things which are \emph{not} presupposed by Bell Locality.  First,
Bell Locality does not assume or require determinism; the condition
is stated in terms of the probabilities assigned to various events
by a given
theory's dynamical laws.  Determinism is of course included as a
special case (in which all the probabilities assume exclusively the
values one or zero), but we do not impose this as an assumption.
Second, the formulation of Bell Locality in no way assumes or requires 
the existence of hidden-variables.  The condition is stated in terms of a
complete state description ($\lambda$) for the particle pair, but
there is nothing in the definition of Bell Locality which requires
$\lambda$ to include anything more than the quantum mechanical wave
function (which, according to orthodox quantum theory, provides 
already a complete description of the state of the particle pair).  
Indeed, it need not even include that.  We make no
assumptions at all about the content of $\lambda$, i.e., we in no way
restrict the class of theories to which we are (initially) open.
Bell Locality simply permits us to easily assess whether a
given theory (of any type, so long as it satisfies the criteria stated
above, i.e., so long as it proposes some definite 
candidate $\lambda$ and permits one to calculate probabilities on 
that basis) is or is not locally causal.

The twist, to be utilized in Section \ref{sec3}, is that we may
infer something about the nature of $\lambda$ from the requirements
that Bell Locality -- and certain empirically-supported facts -- be 
respected.

\section{HIDDEN VARIABLES}
\label{sec3}

Having clarified the definition and meaning of Bell Locality, let us
now turn to the main topic:  What features must a Bell Local theory
possess in order to account for certain empirically-observed facts?  

The empirical fact that will concern us is the following:  if the
particle pair is prepared in (what orthodox quantum theory describes as)
the spin singlet state
\begin{equation}
|\psi\rangle = \frac{1}{\sqrt{2}} \left( |+1\rangle |-1\rangle -
 |-1\rangle |+1\rangle \right)
\end{equation}
and if Alice and Bob happen to choose to measure along the same axis
($\hat{a}=\hat{b}$), the outcomes will be \emph{perfectly anti-correlated}:
either $A = +1$ and $B=-1$, or $A=-1$ and $B=+1$.  Can a Bell Local
theory account for this empirical fact, and if so, how?

To begin, we must explain what is meant by ``preparing the particle
pair in the spin singlet state.''  What this means experimentally and
operationally is clear, but what does it mean in terms of the (as yet
completely unspecified) theory which will (we hope) be able
to account for the results?
Evidently the preparation procedure creates or selects one specific
state out of a class of possible states consistent with the
preparation procedure.  Let us denote this class $ \Lambda_\psi $.
Thus, a particular instance of the preparation will result in 
some particular  $\lambda  \in  \Lambda_\psi $.  
(We consider the set $\Lambda_\psi$ to consist of those
states which can be produced with nonzero probability.  Also note that
we of course leave open the possibility that $\Lambda_\psi$
is a set with only one element, e.g., the wave function itself.
That is, we leave open, for now, the possibility that the 
orthodox quantum state description is \emph{complete}.  Finally, note
that we assume $\Lambda_\psi$ contains a finite number of elements.
Everything that follows can be applied to the case of an infinite 
$\Lambda_\psi$, \emph{mutatis mutandis}.)

Consider now the joint probability for outcomes $A=+1, B=+1$ when
$\hat{a} = \hat{b} = \hat{n}_1$.  In a Bell Local
theory, this joint probability must factor into a product as follows: 
\begin{eqnarray}
&&P(A=+1,B=+1 | \hat{n}_1,\hat{n}_1, \lambda) \nonumber \\
&& \; = P(A=+1 | \hat{n}_1,\lambda) 
\times P(B=+1 | \hat{n}_1,\lambda).
\end{eqnarray}
We are taking it as given (because it is known from experiment) 
that this joint probability vanishes \emph{for
all} $\lambda \in \Lambda_\psi $.  If there were some $\lambda$ that
could, with nonzero probability, be produced by the singlet preparation
procedure, and for which the joint probability above were other than
zero, Alice and Bob would \emph{sometimes} get identical outcomes
even though they measured along the same axis.  Since in fact this 
never happens, we constrain the theories accordingly.

Thus $\forall \; \lambda \in \Lambda_\psi$, either 
\begin{equation}
P(A=+1 | \hat{n}_1, \lambda) = 0
\label{aplus}
\end{equation}
 or 
\begin{equation}
P(B=+1 | \hat{n}_1, \lambda) = 0. 
\label{bplus}
\end{equation}

Now consider the joint probability for outcomes $A = -1$, $B=-1$ when
(again) Alice and Bob both measure along the direction $\hat{n}_1$.
As above, we must have
that $ \forall \; \lambda \in \Lambda_\psi$ either
\begin{equation}
P(A=-1 | \hat{n}_1, \lambda) = 0
\label{aminus}
\end{equation}
or
\begin{equation}
P(B=-1 | \hat{n}_1, \lambda) = 0.
\label{bminus}
\end{equation}
Note that the two conditions -- (\ref{aplus}) \emph{or} (\ref{bplus}), on the
one hand, \emph{and}, on the other, (\ref{aminus}) \emph{or} (\ref{bminus})
-- must be satisfied simultaneously for all $\lambda \in \Lambda_\psi$.  

Further, for any particular $\lambda$, a statement
of the form
\begin{equation}
P(A=+1 | \hat{n}_1, \lambda)=0
\end{equation}
implies one of the form
\begin{equation}
P(A=-1 | \hat{n}_1, \lambda) = 1
\label{aminus1}
\end{equation}
since the outcomes are bivalent:  if, for a given $\lambda$ and a
given measurement direction, a certain outcome is (according to some
theory) \emph{impossible}, 
then, since there are only two possible outcomes, 
the opposite outcome is \emph{required}.  Recall too that the
probabilities here are not empirical relative frequencies, but
probabilities as assigned by the fundamental dynamics of a theory.
The assertion 
here is that if a given theory assigns zero probability to a
certain outcome (under specified conditions), then it must assign
unit probability to the (here only) alternative outcome.  This
suggests a shorthand notation in which we substitute, e.g.,  
for Eq. (\ref{aminus1}) the simpler statement
\begin{equation}
A(\hat{n}_1, \lambda) = -1
\label{aisminus1}
\end{equation}
and similarly for the other possibilities.  

The reader may worry that we are here violating 
Wheeler's famous statement of the orthodox
quantum philosophy:  ``No elementary phenomenon is a phenomenon 
until it is a registered (observed) phenomenon'' \cite{wheelerquote} 
-- i.e., it
is invalid to attribute particular outcomes to experiments which
haven't, in fact, been performed.  This worry is partly justified.
We are not, however,
asserting that an un-performed measurement has an actual, particular
outcome; this would be literal nonsense, and is the grain of truth in
Wheeler's dictum.  Strictly speaking, our statement isn't even
\emph{about} Alice's measurement -- it is about the state $\lambda$
and the theory in which that state assignment is embedded.  The 
real meaning of Eq. (\ref{aisminus1}) is simply this:
for the state $\lambda$, the theory in question
assigns unit probability to the outcome $A = -1$ under the condition
that Alice measures along direction $\hat{n}_1$.  The theory must
attribute sufficient structure to $\lambda$ (and possess the necessary
dynamical laws) such that, should Alice choose to measure along
$\hat{n}_1$, the outcome $A = -1$ is guaranteed.  In this sense, we
may say that the theory in question \emph{encodes} the outcome $A =
-1$ (for measurement along $\hat{n}_1$) in the state $\lambda$.

Crucially, we must not forget that this kind of
statement is simply forced on us by the logical development to this
point:  given that we only consider Bell Local theories, and given
that the theories must be capable of making the correct predictions
for the (possible) case $\hat{a} = \hat{b} = \hat{n}_1$, the theory
\emph{must} posit this sort of structure.  Thus, to whatever extent 
the required theoretical structure conflicts with Wheeler's dictum, 
we must evidently conclude that this orthodox philosophy is excluded 
by our premises (Bell Locality and perfect anti-correlation).

Summarizing the results so far, we must have that, 
$\forall \; \lambda \in \Lambda_\psi$
\begin{equation}
A(\hat{n}_1, \lambda) = -1 \quad {\mathrm{or}} \quad
B(\hat{n}_1,\lambda) = -1
\label{aorb1}
\end{equation}
and
\begin{equation}
A(\hat{n}_1,\lambda) = +1 \quad {\mathrm{or}} \quad
B(\hat{n}_1,\lambda) = +1.
\label{aorb2}
\end{equation}
Since we cannot have both $A(\hat{n}_1,\lambda)=-1$ and 
$A(\hat{n}_1,\lambda)=+1$ for the same $\lambda$ (and likewise for
$B$), Eqs. (\ref{aorb1}) and (\ref{aorb2}) are logically 
equivalent to the following:  either
\begin{equation}
A(\hat{n}_1,\lambda) = -1 \quad {\mathrm{and}} \quad
B(\hat{n}_1,\lambda) = +1
\end{equation}
or
\begin{equation}
A(\hat{n}_1,\lambda) = +1 \quad {\mathrm{and}} \quad
B(\hat{n}_1,\lambda) = -1.
\end{equation}
In other words, the requirements mentioned in Eqs.
(\ref{aplus}) - (\ref{bminus}) will be satisfied if and only if the
states $\lambda \in \Lambda_\psi$ divide into two mutually 
exclusive and jointly exhaustive classes:  those (call them 
$ \Lambda^{  \hat{n}_1 + - }_\psi $) for which
$A(\hat{n}_1,\lambda) = +1$ and $B(\hat{n}_1,\lambda)=-1$ ;
and those (call them $\Lambda^{\hat{n}_1 - + }_\psi $) for which
$A(\hat{n}_1,\lambda) = -1$ and $B(\hat{n}_1,\lambda)=+1$.  
The two classes ( $\Lambda^{\hat{n}_1 + - }_\psi $ and 
$\Lambda^{\hat{n}_1 - + }_\psi  $) are jointly exhaustive
because any $\lambda$ which, with nonzero probability, allows for
identical outcomes on the two sides, will necessarily spoil the
theory's ability to predict perfect anti-correlation.  Any such
$\lambda$ is by definition excluded from $\Lambda_\psi$ to begin
with.  

In a Bell Local theory, the probability for a given outcome on one
side (conditionalized on a complete specification of the state of the
particles) may not depend on goings-on at the distant location.  Given
the empirically observed fact of perfect anti-correlation, we have
shown that a Bell Local theory must attribute
outcome-determining properties to the particle pair.
Putting the same point another way, we have proved that there is no
such thing as a Bell Local theory which accounts for the perfect
anti-correlation without positing deterministic hidden variables
(``hidden'' because these variables evidently go beyond the quantum
state description).  All Bell Local theories which successfully
predict perfect anti-correlation (under the appropriate circumstances)
must posit that the individual
outcomes are determined, in advance of the actual measurements, 
by structure encoded in $\lambda$.  Notice too that the proof that
this kind of theoretical structure is required (for a Bell Local
theory) in no way assumes the \emph{actual} performance of
measurements along $\hat{a} = \hat{b} = \hat{n}_1$.  The mere fact 
that such measurements \emph{may} be later chosen  -- combined with  
the requirement that the theory must be capable of generating the 
appropriate sorts of outcomes should this eventuality arise -- leads
to the conditions described above.

A key point is now the generalizability of this result to measurements
along different axes, say $\hat{n}_2$ and $\hat{n}_3$.  Following
exactly the logic of the above paragraphs, we may argue that
the states $\lambda \in \Lambda_\psi $ must divide into classes
$ \Lambda^{\hat{n}_2 + - }_\psi  $ and 
$ \Lambda^{\hat{n}_2 - + }_\psi  $, such that, respectively,
\begin{equation}
A(\hat{n}_2,\lambda)=+1 \quad {\mathrm{and}} \quad B(\hat{n}_2,\lambda)=-1
\end{equation}
and
\begin{equation}
A(\hat{n}_2,\lambda)=-1 \quad {\mathrm{and}} \quad B(\hat{n}_2,\lambda)=+1.
\end{equation}
And similarly for $\hat{n}_3$.  

Are we, however, going beyond what is required if we insist on
dividing the class $\Lambda_\psi $ using the three measurement
directions simultaneously?  That is, are we permitted to deduce (from
the fact of perfect anti-correlation along \emph{each} of the three
directions $\{ \hat{n}_1, \hat{n}_2, \hat{n}_3 \}$ and the requirement
of Bell Locality) that a theory must posit \emph{eight} mutually
exclusive and jointly exhaustive sub-classes
of $ \Lambda_\psi $ -- for example, the class 
$ \Lambda^{ \hat{n}_1 +-, \hat{n}_2 +-, \hat{n}_3+-}_\psi  $
such that $\forall \; \lambda \in \Lambda^{ \hat{n}_1 +-, \hat{n}_2+-,
  \hat{n}_3+-}_\psi $
\begin{equation}
\left\{
\begin{array}{r@{\; , \quad}l}
A(\hat{n}_1,\lambda) = +1 & B(\hat{n}_1,\lambda)= -1 \\
A(\hat{n}_2,\lambda) = +1 & B(\hat{n}_2,\lambda) = -1 \\
A(\hat{n}_3,\lambda) = +1 & B(\hat{n}_3,\lambda) = -1 
\end{array}
\right\}
\label{oneofeight}
\end{equation}
and similarly for the other seven possible combinations?  This is of
course tantamount to assigning pre-measurement values to the spin
components of each particle along all three considered measurement
directions (and, in principle, along the continuous infinity of
possible measurement directions).  

Such an assignment is obviously in conflict 
with the orthodox quantum philosophy, which expressly
forbids assigning values to unmeasured observables (not to mention
assigning values simultaneously to non-commuting observables!).  
But that is not the relevant issue.  What concerns us here is simply 
this:  is such an assignment required by Bell Locality?  That is, must
a Bell Local theory predicting perfect anti-correlation (along
all of the three considered measurement axes) posit this
detailed structure in the state descriptions?  

The answer is unambiguously yes.  A Bell Local theory (which predicts
successfully the perfect anti-correlation) must posit this structure,
must posit enough detailed content to $\lambda$ that the outcome of
the measurement of any spin component on either particle is
\emph{determined}, as, for example, in Eq. (\ref{oneofeight}).  The
reason is simple:  we are treating the measurement axes $\hat{a}$ and
$\hat{b}$ as \emph{free variables} which can, in principle, be set
randomly (or by a free-will choice) after the pair's state $\lambda$ 
is fixed.  Thus, by precisely the reasoning detailed above, the states
$\lambda \in  \Lambda_\psi $ must be so as to produce perfect
anti-correlation should Alice and Bob happen to both measure along any
of the three considered directions.  This requirement alone does not
necessitate the kind of detailed structure contained in
Eq. (\ref{oneofeight}).  When combined with the requirement of Bell 
Locality, however, this detailed structure -- these deterministic
hidden variables -- \emph{are required}.  A theory (like
orthodox quantum mechanics) which posits less structure for 
$\lambda$ \emph{can}
successfully predict perfect anti-correlation along any of the three
considered directions -- but only at the price of violating Bell Locality.

\section{DISCUSSION}
\label{sec4}

The main result proved in Section \ref{sec3} 
is the following:  All theories
respecting a certain locality condition (Bell Locality) must, in order
to successfully reproduce a certain class of empirically
well-confirmed correlations, posit that the outcomes of all
possible spin-component measurements to be made on the particles are
encoded in the pre-measurement state of the particles, such that the 
outcome in one wing of the experiment is determined, once the
state of the particle pair and the orientation of the nearby apparatus
are specified. \cite{fine} 
In rough terms, the particles must carry ``instruction
sets'' \cite{instructionsets}
which pre-determine the outcomes of spin measurements.  Since
including such ``instruction sets'' in the state specification goes 
beyond what is attributed to the states by orthodox quantum theory,
the kind of theory we have argued for may be termed a hidden-variable
theory.  What is crucial here is that the relevant hidden variables
have not been \emph{assumed}, but rather \emph{derived} -- in
particular, derived from the logical conjunction of (1)
a certain class of empirically supported correlations and 
(2) Bell Locality.  

It is of course well known that a hidden-variable theory of this type
cannot account for another class of empirically tested
correlations, and is therefore not empirically viable.  \cite{aspect}
This is Bell's
Theorem.  It would be a logical fallacy, however, to conclude from
this that the arguments presented in Section \ref{sec3} are flawed.  
This would involve an equivocation between the truth of the \emph{conclusion}
of an argument, and the validity of the argument itself.  What we
have established here is that hidden variables of a particular variety
are required \emph{if one insists on respecting Bell Locality}.  Since
such hidden variable theories are evidently not viable, the proper conclusion
to draw is that Bell Locality cannot be maintained -- i.e., \emph{no theory
respecting Bell Locality can account for the entire class of
empirically observed correlations between distant particles}.  

The logic here was expressed clearly by Bell:
\begin{quote}
``The EPRB correlations are such that the result of the
experiment on one side immediately foretells that on the other,
whenever the analyzers happen to be parallel.  If we do not accept the
intervention on one side as a causal influence on the other, we seem
obliged to admit that the results on both sides are determined in
advance anyway, independently of the intervention on the other side,
by signals from the source and by the local magnet setting.  But 
this has implications for the non-parallel settings which
conflict with those of quantum mechanics.  So we \emph{cannot} dismiss
intervention on one side as a causal influence on the other.''
\cite[pg 149-50]{bell}
\end{quote}
The same basic argument has also been advocated by a number of other
authors, for example those in the first
of the ``camps'' described in Section \ref{sec1}.  The present paper adds two
notable features:  first, a precise mathematical clarification of 
Bell's phrase ``if we do not accept the intervention on one side as 
a causal influence on the other'' and, second, a rigorous mathematical
derivation of what Bell merely says ``we seem obliged to admit.''  

H. P. Stapp apparently agrees with Bell that one ``cannot dismiss
intervention on one side as a causal influence on the other.''
But, as discussed in the introduction, Stapp's argument
for this conclusion is different from that advocated by Bell.
And since Stapp's arguments have been shown to suffer
from several apparently fatal flaws, it is worth clarifying in some
detail how Bell's argument (elaborated and sharpened in the present
paper) differs from Stapp's.

First, let us emphasize the point about overall logical structure.  
Stapp has attempted to prove a
``Bell-type theorem without hidden variables'' -- i.e., to prove a
\emph{stronger} version of Bell's Theorem which dispenses with one
of the premises on which Bell's own derivation rests.  By contrast,
we retain the original (``weak'') version of Bell's Theorem, and
\emph{supplement} it by showing that the two premises used in deriving the
inequality -- (i) locality and (ii) hidden variables -- are not
logically unrelated axioms.  Rather, premise (ii) follows from
premise (i) and the empirical fact of perfect anti-correlation.

The criticisms made against Stapp by Mermin, Unruh, Shimony, Stein,
and others suggest that Bell's Theorem cannot be
strengthened in the way that would be required to validate Stapp's
route to our shared conclusion.  One simply cannot arrive at a Bell
Inequality (or the corresponding logical contradiction arrived at in
the context of Hardy's Theorem) if one rules out any assumption of
hidden-variables or counterfactual-definiteness and thus
restricts the discussion to actual measurement results.

The reason such a restriction precludes a valid non-locality proof
is straightforward.  In any particular run
of the experiment, only one measurement can actually be performed in each
wing.  There are, thus, two and only two actual measurement outcomes
to work with.  But any such pair of outcomes can \emph{always}
and \emph{trivially} be accounted for locally -- namely, by simply
asserting that the measurements revealed locally pre-existing values
for the observables in question.  Of course, one is then no
longer talking about orthodox quantum theory but, rather, about some
kind of local hidden-variable theory.  But if one has decided to focus
exclusively on measurement outcomes (and not on the specific theories
which predict those outcomes), this is a distinction without a
difference and no nonlocality can be established.

One might also allow talk of the relative frequencies with which
various outcomes appear.  It would then be possible to define
locality by the insistence that the
relative frequency of a given outcome on one side (say, $B$) 
not correlate with the freely chosen setting ($\hat{a}$) on the other side.  
\footnote{Of course, one could also ask whether the relative frequency 
of a given outcome on one side ($B$) correlates with the
\emph{outcome} ($A$) on the far side.  But, as explained in Section
\ref{sec2}, such a correlation does not necessarily entail any 
nonlocal causation.  Thus again, as long as one arbitrarily restricts
one's attention to measurement results rather than the structure of
the theories which predict those results, nonlocal causation cannot 
be proved (or even, really, clearly defined).  On a related point, 
this seems to be the reason that otherwise sensible people came to
accept that ``parameter dependence'' was relativistically excluded,
while ``outcome dependence'' was consistent with relativity.  If one
identifies probabilities with \emph{relative frequencies}, a violation
of ``parameter independence'' (plus some other conditions) 
permits superluminal signaling (which
is clearly outlawed by relativity) while a violation of ``outcome
independence'' does not.  But if one is speaking not of relative 
frequencies but of the fundamental dynamics of a theory, then any
dependence of probabilities on space-like separated events should
obviously be excluded on relativistic grounds.  This is why Bell
Locality (as opposed to any of the various weaker conditions into
which it can in principle be analyzed) is such a reasonable
formalization of relativity's prohibition on superluminal causation.}  
In our terminology, this condition would read
\begin{equation}
P(B|\hat{a},\hat{b}) = P(B|\hat{a}',\hat{b})
\end{equation}
where 
\begin{equation}
P(B|\hat{a},\hat{b}) = \sum_{\lambda \in \Lambda_\psi }
\sum_{A = \pm 1} P(A,B|\hat{a},\hat{b},\lambda) \; P_\psi(\lambda),
\end{equation}
$\hat{a}$ and $\hat{a}'$ are two different possible 
(freely-choosable) settings of the apparatus on the left, 
and $P_\psi(\lambda)$
is the probability that the preparation procedure produces the
particular state $\lambda$.
This condition is called Signal Locality because its
violation would permit Alice to send superluminal signals to Bob.
It is well known that orthodox quantum theory does not violate Signal
Locality. \cite{ballentinejarrett}
Moreover, since the condition refers only to relative
frequencies of outcomes, all theories which share the same empirical
predictions as orthodox quantum theory (for example, Bohmian
Mechanics) also respect Signal Locality.  

Either way, so long as one restricts one's analysis 
to exclusively observable phenomena, there is no hope of 
establishing nonlocality.
And this is the deepest reason that Stapp's project fails.  In his
original paper, Stapp defines locality as follows:  ``no free choice 
can influence observable phenomena lying outside its forward light 
cone''. \cite{stapp1}  But for the reasons just indicated, this
definition (which is apparently equivalent to Signal Locality) is
too restrictive.  It is impossible to establish the reality of a violation of
this sort of locality.  

Stapp attempts to get around this difficulty
by using, in his actual proof, a less restrictive definition of locality.
(Thus, he regards the fact that the truth of his statement $S$
depends on a distant free choice, as establishing a violation of
locality even though $S$ -- a counterfactual conditional -- is not an
``observable phenomenon''.)  And this stronger definition of locality
is precisely what his critics have criticized:  the stronger
definition ``can be interpreted as assigning an unwarranted level of
reality to the value of certain quantum attributes.''  \cite{unruh}
That is, it doesn't comport with the positivist or phenomenalist
approach of the orthodox quantum philosophy.

Stapp repeatedly stresses that ``all the 
assumptions used in [his] proof are elements of orthodox quantum 
philosophy''. \cite{stappresponse}  At root, it is precisely Stapp's
allegiance to this philosophy which prevents him from constructing a
valid proof.  The conclusion he wants to establish (which is about
\emph{causality}, not merely the ability to send signals) is,
according to that positivist philosophy, off the table 
from the very beginning.

The phenomenalist attitude of the orthodox quantum philosophy also 
invites the identification of quantum \emph{theory} with any theory 
sharing its empirical \emph{predictions}.  
\footnote{This conflation already arose above,
  when it was pointed out that the ``obvious and trivial'' local
  explanation for a given pair of outcomes involves a rejection of the
  orthodox completeness doctrine.  On the phenomenalist premise,
  however, the orthodox theory and the local hidden-variable theory
  are literally the same theory, and it becomes impossible to sort out
  which version is and which version isn't local.  Note also that this
  phenomenalist identification of a theory with its empirical
  predictions renders the orthodox completeness claim literally
  meaningless.}
Stapp's acceptance of this philosophy then leads to a recurring confusion 
between two different goals:  showing that quantum theory
itself is nonlocal, and showing that any theory sharing quantum
theory's empirical predictions must be nonlocal.  For example, despite 
apparently attempting to show 
that some sort of nonlocality is required by the quantum mechanical 
predictions (i.e., required of any theory which shares those 
predictions), Stapp titles his paper ``Nonlocal character of quantum
theory.''  

In fact, though, establishing the nonlocal character of
orthodox quantum theory is \emph{easy}:  once (an appropriate
sense of) locality is defined precisely (as in Section \ref{sec2}), 
it is a trivial observation that the
orthodox theory (where $\lambda$ is simply $\psi$)
fails to respect Eq. (\ref{factorize}) and 
hence is not local.  The \emph{difficult} assignment is to
show not just that orthodox quantum theory is nonlocal, but that this
nonlocality is ineliminable -- that no Bell Local theory can share the
quantum mechanical predictions (or, equivalently, can match experiment).  
Stapp's allegiance to the orthodox quantum philosophy thus ensures in 
advance that his project cannot succeed: when he inevitably slides over
from the easy assignment to the hard assignment, he necessarily goes
beyond standard quantum theory and thus leaves himself open to the
objection that he is ``assigning an unwarranted level of reality to''
certain things.  Given his expressed premises, this objection is
unavoidable and completely fatal.

Stapp once described his project as follows:
``The nonlocality that I claim to exhibit is
completely compatible with the locality properties of relativity
theory, which, in a quantum context, pertain only to features of
\emph{our observations}, not to features of a putative underlying
reality.'' \cite{stappresponse}  
Since the empirical predictions of quantum 
theory respect Signal Locality, there is no way to ``exhibit''
any nonlocality at the level of ``our observations.''  It simply
cannot be done.  But if, motivated by the orthodox quantum philosophy,
one excludes from the beginning any talk about the ``features of a
putative underlying reality'', then there is literally nothing else --
that is, no other sense of locality -- to discuss.  
The vague anti-realism of 
the orthodox quantum philosophy thus seems to rule out the 
very kind of talk that is absolutely required to show that nature 
violates some locality condition -- namely, talk of nature!  
But orthodox quantum theory better commit to a realistic
description of \emph{something}.  Otherwise -- that is, if one retreats
to an exclusively epistemological or ``algorithmic'' 
interpretation of quantum theory in
general and the wave function in particular -- one simply no longer 
has a \emph{theory} in the sense defined in Section \ref{sec2}.  It
is then \emph{meaningless} to discuss whether the causal
processes posited by the ``theory'' respect relativity's prohibitions
on superluminal causation.  A formalism which is not \emph{about} any
such processes
is neither local nor nonlocal.  Both terms are simply inapplicable.
\cite{boxes}

One can start to see the shiftiness of the orthodox quantum 
philosophy which led Bell to describe it as ``unprofessionally vague
and ambiguous.''\cite[pg 173]{bell}  Ultimately, it is Stapp's 
acceptance of this philosophy which undercuts his attempt to 
exhibit nonlocality.  Specifically, it prevents him (a) from
articulating a clear definition of local causality, (b) from clearly 
noting the nonlocality that is already present in the orthodox 
theory, and (c) from even considering the types of theories that
might have been thought to have provided a local explanation of
the correlations in question.

In the alternative to Stapp's approach advocated by Bell and 
elaborated here, we begin not with an 
arbitrary allegiance to the orthodox quantum philosophy, 
but with a straightforward, mathematically precise definition of
local causality, motivated exclusively by relativity theory.
We then ask:  is there a theory which respects this
locality condition and which is consistent with what we know from
experiment about the correlations between distant particles (prepared
in a certain way)?  The answer turns out to be no.  To see this, it
is helpful to break the question into two stages:
\begin{enumerate}

\item  \emph{What structure must a theory have if it is to respect the
locality condition and successfully match a certain set of
empirical facts?}  The precise answer to this question was provided
in Section \ref{sec3} above.

\item  \emph{Is a theory with this structure consistent with all of
    the other empirical facts?}  The answer to this question is given
    by Bell's Theorem in its original version:  no.

\end{enumerate}
The main strength of this two-step approach is simply that it permits an 
interesting, yet unambiguous, conclusion:  No Bell Local theory can be
empirically viable.  

Of course, one could always
assert that the conclusion isn't, appearances to the contrary
notwithstanding, all that interesting, since Bell Locality isn't
a correct formalization of relativity's
prohibition on superluminal causation.  One possible
argument for this view would be that orthodox quantum theory 
violates Bell
Locality:  if one insists on remaining faithful to the orthodox
philosophy (and ensuring that quantum theory 
doesn't turn out to be nonlocal)
then Bell Locality \emph{can't} be the ``right'' definition of
locality.

That argument is too obviously question-begging 
to deserve attention.  
Is there some more reasonable objection that could be given against
the apparent appropriateness of Bell Locality?  Perhaps.  But
Bell's case -- elaborated in Section \ref{sec2} above -- is, on its face,
sufficiently plausible that the burden of proof should lie with those
who reject this locality condition.  If Bell Locality isn't a good
formalization of ``relativistic causality'', why not?  And what is?

Until and unless such questions can be answered, we must evidently
agree with Bell that there exists an ``incompatibility, at the deepest
level, between the two fundamental pillars of contemporary theory...''
\cite[pg 172]{bell}  No Bell Local theory can account for the
empirically verified predictions of quantum theory.  Nature
is not Bell Local.  

And, to return to the question of the hidden
variables program with which we began, this surely means that the
use of Bell's Theorem against Bohmian Mechanics (by the ``second 
camp'') is fundamentally misguided.  Orthodox quantum theory
itself violates Bell Locality, so the fact that hidden variable
theories like Bohm's violate it as well is no argument in support
of the standard theory.  There simply is no Bell Local theory that is
in agreement with experiment, so it is ridiculous to reject any one
particular theory (which does agree with experiment!) merely on the
grounds that it violates Bell Locality.
Even leaving aside
the question of the appropriateness of Bell Locality (as a
formalization of relativistic causality), it is an inescapable (just,
someone might conceivably argue, uninteresting) fact that the 
failure of Bell Locality is a 
real feature of the world.  So it can hardly be a valid objection 
against Bohmian Mechanics (or any other theory) 
that it displays this feature.


\begin{thebibliography}{00}

\bibitem{bell}
J. S. Bell, \emph{Speakable and Unspeakable in Quantum Mechanics} 
(Second Edition), (Cambridge University Press, Cambridge (UK), 2004)

\bibitem{stappquote}
H. P. Stapp, ``Bell's Theorem and World Process,'' \emph{Nuovo
  Cimento} {\bf{29B}}, 270-6, (1975)

\bibitem{maudlin}  T. Maudlin, \emph{Quantum Non-Locality and
    Relativity} (Second Edition), (Blackwell, Cambridge (MA), 2002)

\bibitem{shelly}  D. D\"urr, N. Zanghi, and S. Goldstein, ``Quantum
  Equilibrium and the Role of Operators as Observables in Quantum
Theory,'' \emph{Journal of Statistical Physics} {\bf{116}}, 959-1055,
  (2004).  (See in particular Section 8:  Hidden Variables.)

\bibitem{norsen}  T. Norsen, ``EPR and Bell Locality,''
  quant-ph/0408105, to appear in \emph{Are there Quantum Jumps? and On
  the Present Status of Quantum Mechanics}, A. Bassi, D. D\"urr,
  T. Weber, and N. Zanghi, \emph{eds.}, AIP Conference Proceedings, 2006

\bibitem{wiseman}  H. M. Wiseman, ``From Einstein's Theorem to Bell's 
Theorem:  A History of Quantum Nonlocality,'' \emph{Contemporary
  Physics} {\bf{47}}, 79-88 (2006)

\bibitem{wigner}  E. Wigner, ``Interpretation of
  Quantum Mechanics,'' (1976), reprinted in \emph{Quantum Theory and
  Measurement}, J. A. Wheeler and W. H. Zurek, \emph{eds.}
  (Princeton University Press, Princeton, 1983)  

\bibitem{merminquote}  N. David Mermin, ``Hidden
  Variables and the Two Theorems of John Bell,'' \emph{Reviews of Modern
  Physics} {\bf{65}}, 803-815, (July 1993)

\bibitem{epr}  Einstein, Podolsky, and Rosen, ``Can quantum-
  mechanical description of physical reality be considered
  complete?'' \emph{Physical Review} {\bf{47}}, 777-780, (1935)

\bibitem{bohmmech}  S. Goldstein, ``Bohmian Mechanics,'' \emph{The
    Stanford Encyclopedia of Philosophy}, Edward
    N. Zalta (ed.), http://plato.stanford.edu/entries/qm-bohm; see
    also R. Tumulka, ``Understanding Bohmian mechanics: a dialogue,''
    \emph{American Journal of Physics} {\bf{72}}, 1220-6, (2004)

\bibitem{stapp:btwhv}  H. P. Stapp, ``Bell's Theorem Without Hidden
  Variables,'' quant-ph/0010047

\bibitem{stapp1}  H. P. Stapp, ``Nonlocal character of quantum
  theory,'' \emph{American Journal of Physics} {\bf{65}}, 300-304, (1997) 

\bibitem{stapp2}  H. P. Stapp, ``A Bell-type theorem without hidden
  variables,'' \emph{American Journal of Physics} {\bf{72}}, 30-33, (2004)

\bibitem{hardy}  Lucien Hardy, ``Quantum mechanics, local realistic
  theories, and Lorentz invariant realistic theories,''
  \emph{Physical Review Letters} {\bf{68}}, 2981-2984, (1992)

\bibitem{unruh} W. Unruh, ``Is Quantum Mechanics Non-Local?''
  \emph{Physical Review A} {\bf{59}}, 126-130, (1999)

\bibitem{shimony} A. Shimony, ``An Analysis of Stapp's `A Bell-type
  theorem without hidden variables','' quant-ph/0404121

\bibitem{shimonystein}  A. Shimony and H. Stein,
  ``Comment on `Nonlocal character of quantum theory,'...,''
  \emph{American Journal of Physics} {\bf{69}}, 848-853, (2001)
 
\bibitem{mermin}  N. David Mermin, ``Nonlocal character of quantum
  theory?'' \emph{American Journal of Physics} {\bf{66}}, 920-4, (1998)

\bibitem{mermin2}  N. David Mermin, ``Nonlocality and Bohr's reply to
  EPR,'' quant-ph/9712003

\bibitem{bellwohv} H. P. Stapp, ``Bell's Theorem Without Hidden
  Variables,'' \emph{op cit.}; See also P. Eberhard, ``Bell's Theorem
  Without Hidden Variables,'' \emph{Nuovo Cimento} {\bf{38B}}, 75-80, (1977)

\bibitem{eberhard}  P. Eberhard, ``Bell's Theorem and the different
  concepts of locality,'' \emph{Nuovo Cimento} {\bf{46B}}, 392-419, (1978)

\bibitem{wheelerquote}  J. A. Wheeler, ``Law Without Law'' in
  J. A. Wheeler and W. H. Zurek, \emph{eds.}, \emph{Quantum Theory and
  Measurement}  (Princeton University Press, Princeton, 1983)

\bibitem{fine} Thanks to Arthur Fine for, after reading an initial
  draft of the current paper, pointing out to me that similar arguments
  have appeared previously in the literature:  e.g., Brian Skyrms,
  ``Counterfactual Definiteness and Local Causation,'' \emph{Philosophy of
  Science} {\bf{49}}, 43-50, (1982); Patrick Suppes, ``Some remarks on hidden
  variables and the EPR paradox,'' \emph{Erkenntnis} {\bf{16}}, 
  311-314, (1981) (and references therein).  

\bibitem{instructionsets}  N. D. Mermin, ``Bringing home the atomic
  world: Quantum mysteries for anybody,'' \emph{American Journal of
  Physics} {\bf{49}}, 940-943, (1981)

\bibitem{aspect} G. Weihs, T. Jennewein, C. Simon, H. Weinfurter, and
  A. Zeilinger, ``Violation of Bell's inequality under strict Einstein
  locality conditions,'' \emph{Physical Review Letters} {\bf{81}},
  5039-5043, (1998)

\bibitem{ballentinejarrett} L. Ballentine and J. Jarrett, ``Bell's
  theorem: does quantum mechanics contradict relativity?''
  \emph{American Journal of Physics} {\bf{55}}, 696-701, (1987)

\bibitem{stappresponse}  H. P. Stapp, ``Response to `Comment on
  ``Nonlocal character of quantum theory'' ' by Abner Shimony and 
  Howard Stein ...,'' \emph{American Journal of Physics} {\bf{69}}, 
  854-9, (2001)

\bibitem{boxes} For some additional discussion of this point, see:  
 T. Norsen, ``Einstein's Boxes,''
  \emph{American Journal of Physics} {\bf{73}}, 164-176, (2005)



\end{thebibliography}
\end{document}